# Quantum Key Distribution With several intercept-resend attacks Via A Depolarizing Channel

Mustapha Dehmani, Mohamed Errahmani, Hamid Ez-Zahraouy* and Abdelilah Benyoussef

LMPHE (URAC 12), Faculté des Sciences, Université Mohammed V-Agdal, B.P.1014, Rabat, Morocco

*Corresponding author: ezahamid@fsr.ac.ma

**Abstract**

The disturbance effect of a depolarizing channel on the security of the quantum key distribution of the four state BB84 protocol, with multiple sequential intercept-resend attacks of many eavesdroppers, has been studied. The quantum bit error rate and the mutual information are computed for an arbitrary number N of eavesdropper. It is found that the quantum error rate decreases when increasing the depolarizing parameter p characterizing the noise of the channel. For p<0.165, there exist, a special value $p_{tr}$ of p below which the information is secured, otherwise it is not secured. The value of $p_{tr}$ decreases when increasing the number of attacks. While, for p ≥ 0.165, the information is not secured independently of the eavesdropper's number. Phase diagrams corresponding to the secured-unsecured information are also established.

# I-Introduction

The ways of cryptography and quantum data processing crossed when Charles Bennett and Gilles Brassard developed the first quantum key distribution (QKD) known as BB84 protocol [1]. This protocol allows two participants to exchange a random secret key that can be used perfectly for protected communications. Many researchers proposed formal evidences of the QKD safety [2-4]. The QKD was already made in practice, and the first prototype was developed in 1989 by Bennett et al. [5-7]. In spite of experimental successes, several obstacles are important regarding the deployment of this technology out of laboratories [8]. One of these factors is the too short distances reached by prototypes; another factor is the depolarizing channel and its effect with the presence of eavesdroppers [9]. In a previous work, we have analyzed the quantum key distribution with several intercept and resend attacks [10] and with several cloning attacks [11]. Recently, the most important problem in practical quantum communication is the security of the information under the effect of disturbances or quantum noise. In general, the losses of information may be due to several factors namely; the attenuation of photon signal due to the weak pulse laser [12], the detector efficiency mismatch due to the high dark counts [13] and the the depolarizing channel noise effect [14]. In the two former cases the amount of the losses information, in the absence of eavesdropping, depends strongly on the length of the channel. Consequently, one can easily study the dependence of the quantum bit error rate and the mutual information as a function of the length of the channel, even, in the presence of eavesdroppers.

However, in the case of Pauli channel [14], the disturbance depends only on the depolarizing parameter p, and on the length of the channel [15,16]. Recently, it has been studied in different point of view, theoretical and experimental [12,15–18].

Our aim in this paper is to study the security of the information of the four state BB84 protocol, under the effect of both channel noise and many eavesdroppers with intercept-resend attacks.

The paper is organized as follows: The protocol is detailed in section II. Section III is devoted to results and discussion, while section IV is reserved to conclusion.

# II The Protocol

The BB84 protocol requires two different phases: the first one is through the quantum channel physics with one-way, and the second one is through a bidirectional traditional ideal channel authenticated [1]. The quantum depolarization channel, we will study in this paper, leaves intact a qubit $|\psi\rangle$ with probability (1-p) or applies one of the Pauli matrices $\sigma_i$ with probability p/3 for each one [14, 15], which is mathematically considered as an operator $T_p$ defined by:

$$T_p(|\psi\rangle\langle\psi|) = (1-p)|\psi\rangle\langle\psi| + \sum_{i=1}^{3} \frac{p}{3}\sigma_i|\psi\rangle\langle\psi|\sigma_i \quad (1)$$

We denote by $P^*(x_B/x_A)$ the conditional probability that Bob receives a polarized photon $x_B=0,1$ with respect that Alice sends a polarized photon $x_A=0,1$ under the channel noise effect, namely:

$$P^*(0/0) = P^*(1/1) = 1 - \frac{2p}{3} \quad (2)$$

and

$$P^*(1/0) = P^*(0/1) = \frac{2p}{3} \quad (3)$$

Equations (2) and (3) show that, in the absence of eavesdroppers, the information safety depends only on the depolarizing channel parameter p with a quantum bit error $\delta = \frac{2p}{3}$.

In the presence of one eavesdropper within intercept and resend attack, it is difficult to know on which side of the channel the Pauli transformations, corresponding to the channel noise, take place? Between Alice and Eve or between Eve and Bob; In the other side, the model we used does not allow us to know with certainty whether the photon sent by Alice is disturbed, under the channel noise, before or after being intercepted by Eve. Physically, these two situations are completely different, and lead to different results. For example, in the case of one eavesdropper, the two situations are considered but with different probabilities q and 1-q respectively as has been used in reference [9]

Moreover, in the particular cases, for which the position of Eve is closer to Alice and the position of Eve is closer to Bob, we consider q=0 and q=1 respectively.

In the case of many sequential intercept-resend attacks, a state sent by Alice can follow the ways of transmission according to Fig.1, where the photon losses its polarization, due to the channel noise, in one of the space intervals located between Alice and $E_1$, or $E_1$ and $E_2$, ..., or $E_i$ and $E_{i+1}$, ..., or $E_n$ and Bob, with the respective probabilities $q_1, q_2, ... q_i, ..., q_{n+1}$ and $\sum_{i=1}^{n+1} q_i = 1$,

Alice sends a sequence of photons to Bob by choosing randomly to send 1 or 0. Bob chooses randomly to measure the received photon. Between them, there are N independent eavesdroppers $E_i$ (i=1,...,N) who intercept sequentially the photon sent by Alice with the respective probabilities $\omega_i$. The eavesdropper $E_1$ intercepts the photon sent by Alice with the probability $\omega_1$, measures its polarization and sent it to $E_2$, then each eavesdropper $E_i$ intercepts with the probability $\omega_i$ the photon sent by $E_{i-1}$, measures its polarization by choosing randomly a base, and return them to $E_{i+1}$, in the state of polarization that she has measured and son on…, until the eavesdropper $E_N$ who sent its intercepted photon to Bob. However, in the place of the photon that they didn't measure, they put randomly 0 or 1 in their chains of bits. Then, Alice and Bob exchange in a traditional way the bases that they used; they remove in their chain of bits those exchanged in different bases. For studying the security of information exchanged between two honest parties Alice and Bob, we introduce the notion of mutual information. In this case we compute the mutual information between Alice and Bob and the mutual information between Alice and every eavesdropper $E_m$.

**III-Results and discussion**

Moreover it is demonstrated that the binary Shannon entropy allows finding the impact on the information safety [19-21]. However, the mutual information I(A,B) and I(A,$E_m$) between Alice and Bob, and between Alice and the m$^{th}$ eavesdropper $E_m$, are given respectively as follows:

$$I(A,B)=1+P_{AB}(0/0)Log_2(P_{AB}(0/0))+P_{AB}(1/0)Log_2(P_{AB}(1/0)) \qquad (4)$$

$$I(A,E_m)=1+P_{AE_m}(0/0)Log_2(P_{AE_m}(0/0))+P_{AE_m}(1/0)Log_2(P_{AE_m}(1/0)) \qquad (5)$$

Where the conditional probabilities $P(x_B/x_A)$ and $P(x_{E_m}/x_A)$ between Alice and Bob and Alice and the m$^{th}$ eavesdropper. Before the establishment of the general expressions of these probabilities to an arbitrary eavesdropper number N, let us consider the simple case of N=1. In this case, Alice sends a polarized photon $|0\rangle$ or $|1\rangle$ to Bob. This photon can undergo a depolarization channel according to the eq.(1), between Alice and Eve with a probability $q_1$ or between Alice and Eve with a probability $1-q_1$.

Hence, we denote hereafter by $P^0{}_{AB}(x_B/x_A)$ and $P^0{}_{AE}(x_E/x_A)$ the conditional probabilities, in the case of a perfect channel (p=0), that Bob (Eve) receives a polarized photon $x_B(x_E)$=0,1 with respect that Alice sends a polarized photon $x_A$=0,1 namely: $P^0{}_{AB}(0/0)=1-\frac{\omega}{4}$ and $P^0{}_{AE}(0/0)=\frac{1}{2}+\frac{\omega}{4}$. However, using equations (2) and (3), the conditional probabilities $P_{AB}(x_B/x_A)$ between Alice and Bob, under the effect of both channel noise and eavesdropper attacks, are given as follows:

$$P_{AB}(0/0) = P^*(0/0)P^0{}_{AB}(0/0) + P^*(0/1)P^0{}_{AB}(0/1) = (1 - \frac{2p}{3})\left(1 - \frac{\omega}{4}\right) + \frac{\omega p}{6} \qquad (6)$$

$$P_{AE}(0/0) = \left(P^0{}_{AE}(0/0)P^*(0/0) + P^0{}_{AE}(0/1)P^*(0/1)\right)q_1 + P^0{}_{AE}(0/0)(1 - q_1) = \frac{1}{2} + \frac{\omega}{4} - \frac{p\omega q_1}{3} \qquad (7)$$

With, $P_{AB}(0/1) = P_{AB}(1/0) = 1 - P_{AB}(0/0)$, $P_{AB}(1/1) = P_{AB}(0/0)$, $P_{AE}(1/1) = P_{AE}(0/0)$ and $P_{AE}(0/1) = P_{AE}(1/0) = 1 - P_{AE}(0/0)$

The above results can be generalized to the case of an arbitrary number N of eavesdropper as follows:

$$P_{AB}(0/0) = \left(\sum_{k=0}^{n} \frac{2^{n-k}+1}{2^{n-k+1}} \sum_{i_1,\ldots,i_k=1,n} \prod_{j=1}^{k}(1-\omega_{i_j}) \prod_{l=k+1}^{n} \omega_{i_l}\right)\left(1 - \frac{4p}{3}\right) + \frac{2p}{3} \qquad (8)$$

$$P_{AE_m}(0/0) = \left(\left(1 - \frac{4}{3}p\right)\sum_{k=0}^{m-1} \frac{2^{m-k}+1}{2^{m-k+1}} \sum_{i_1,\ldots,i_k=1,m-1} \prod_{j=1}^{k}(1-\omega_{i_j}) \prod_{l=k+1}^{m} \omega_{i_l} + \frac{1-\omega_m}{2} + \frac{2}{3}p\omega_m\right)\sum_{i=1}^{m} q_i$$
$$+ \left(\frac{1-\omega_m}{2} + \sum_{k=0}^{m-1} \frac{2^{m-k}+1}{2^{m-k+1}} \sum_{i_1,\ldots,i_k=1,m-1} \prod_{j=1}^{k}(1-\omega_{i_j}) \prod_{l=k+1}^{m} \omega_{i_l}\right)\left(1 - \sum_{i=1}^{m} q_i\right) \qquad (9)$$

With $P_{AB}(1/0) = P_{AB}(0/1) = 1 - P_{AB}(0/0)$, $P_{AB}(1/1) = P_{AB}(0/0)$, $P_{AEm}(1/0) = P_{AEm}(0/1) = 1 - P_{AEm}(0/0)$ and $P_{AEm}(1/1) = P_{AEm}(0/0)$

Finally, the lost information between Alice and Bob is given by:

$$I_{lost} = I(A,E) + H(\delta) \qquad (10)$$

Where, I(A,E) corresponds to the maximum of the mutual information intercepted by all the eavesdroppers, namely:

$$I(A,E) = \underset{m=1,N}{Max}[I(A,E_m)] \qquad (11)$$

and $H(\delta) = -(1-\delta)Log_2(1-\delta) - \delta Log_2 \delta$ is the binary Shannon entropy, corresponding to the amount of the information lost under the channel noise, in the absence of any external eavesdropper attacks. Hence, the error probability $P_{err}$ is given by [21,22]:

$$P_{err} = \sum_{x_A \neq x_B} [P_{AB}(x_A, x_B)|_{\substack{p \neq 0 \\ \omega_i \neq 0}} - P_{AB}(x_A, x_B)|_{\substack{p=0 \\ \omega_i = 0}}] \qquad (12)$$

The quantum error QBER, due to both channel noise and external attacks, is the value of the error probability $P_{err}$ for which [23-25]

$$I(A,B) = I(A,E) + H(\delta) \qquad (13)$$

For $P_{err}<Q_{BER}$, $I(A,E)+H(\delta)<I(A,B)$, then the information is secured, while it is not secured for $P_{err}>Q_{BER}$ for which $I(A,E)+H(\delta)>I(A,B)$

Equation (8) shows that the amount of the information received by Bob is independent of the stochastic parameter q governing on which region in the channel the depolarization of the photon occurs. While equation (9) shows, that the amount of the intercepted information by any eavesdropper $E_m$ depends strongly on the regions located between Alice and $E_m$, where the photon can loss its polarization with different probabilities $q_1$, $q_2$,..., $q_m$. Hence, the mutual information are computed numerically for N=1, 2 and 3, where the eavesdropper attack probabilities $\omega_i$ are independent. For N=1, the behavior of the quantum bit error QBER as a function of the depolarizing parameter p is shown in Fig.2 a. It is clear that the quantum error decreases when increasing p and depends also on the coefficient $q_1$=q which means the dependency of the amount of the intercepted information by Eve on his position in the channel, i.e, the intercepted information depends on which the depolarization of the photon occurs between Alice and Eve or between Eve and Bob (Fig.2b). It is clear that the lost information increases when decreasing q, which means, in other word, that the amount of information intercepted by Eve becomes important, when the depolarization of the photon occurs between Eve and bob and not between Alice and Eve. Beside the lost information increases when increasing the depolarizing parameter p, passes through a maximum and decreases for sufficiently large value of p. Phase diagram showing the region in the ($p$,$\omega$) plane where the information is secured, is given in Fig.3. It is found that, for a fixed value of the attack probability $\omega$, there exist a special value of p=$p_{tr}$, at which the secured-unsecured transition occurs. Hence, for p<$p_{tr}$, the amount of the lost information is less than that received by Bob, which means that the information is secured. While for p>$p_{tr}$, the information is not secured. To better understand the influence of sequential attacks, by three eavesdroppers, on the security of the information, we have established numerically, the phase diagram in the space parameter ($\omega_1$,$\omega_2$,$\omega_3$) showing the transition between secured-unsecured information for $q_i$=1/4, p=0.05 (Fig.4a) and p=0.1 (Fig.4b). It is clear that the transition surface between secured and secured information depends strongly on the values of p and the secured area decreases when increasing the p.

Now we investigate the security phase diagram (Fig.5) and the behavior of the quantum bit error (Fig.6) in the particular case corresponding to an arbitrary number of eavesdroppers intercepting the photon with identical probabilities $\omega_i = \omega$, but without sharing the results of their measures. Phase diagram established in Fig.5 shows that the secured area decreases when increasing the number of eavesdropper since the amount of intercepted information increases with N. On the other hand, it is clear from Fig.6 that the quantum error depends strongly on the number of eavesdropper for small depolarizing parameter values, while this dependence is weak for sufficiently large values of p, especially near p=0.165.

**IV-Conclusion:**

We have studied the effect of both channel noise and many sequential intercept-resend attacks on the security of information in the BB84 protocol. We have shown that the quantum error decreases when increasing the number of sequential attacks and/or the depolarizing parameter. For a fixed depolarizing parameter p, the security of information becomes weak when increasing the number of eavesdropper, which leads to a secured-unsecured transition at a special number of attacks. This number decreases when increasing p.

**Figure Captions**

**Fig.1**: Scheme of the quantum channel with many sequential attacks

**Fig. 2**: The behaviour of the quantum bit error (a) and the lost information (b) as a function of the depolarizing parameter $p$ for $N=1$ and several values of q.

**Fig. 3**: Phase diagram in the $(p,\omega)$ plane showing the transition between secured and unsecured information for $N=1$ and $q_1=1/2$

**Fig. 4**: Phase diagram in the space parameter $(\omega_1,\omega_2,\omega_3)$ showing the transition between secured and unsecured information for $N=3$ and $q_1=q_2=q_3=q_4=1/4$ with (a) p=0.05 and (b) p=0.1

**Fig. 5**: Phase diagram in the $(p,\omega)$ plane showing the secured-unsecured transition in the case of $\omega_i=\omega$ (i=1,..., N) and $q_i=q=1/(N+1)$ and different values of N.

**Fig.6**: The behaviour of the quantum bit error as a function of the depolarizing parameter $p$ for $\omega_i=\omega$ (i=1,..., N) and $q_i=q=1/(N+1)$ and several values of N

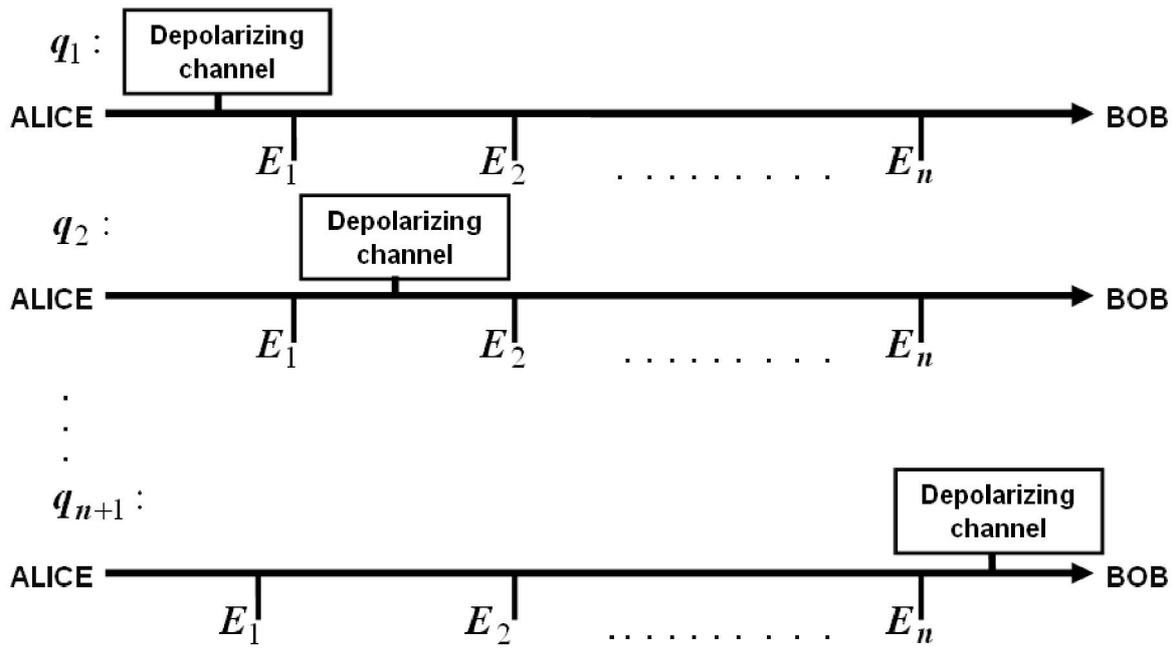

**Fig.1**

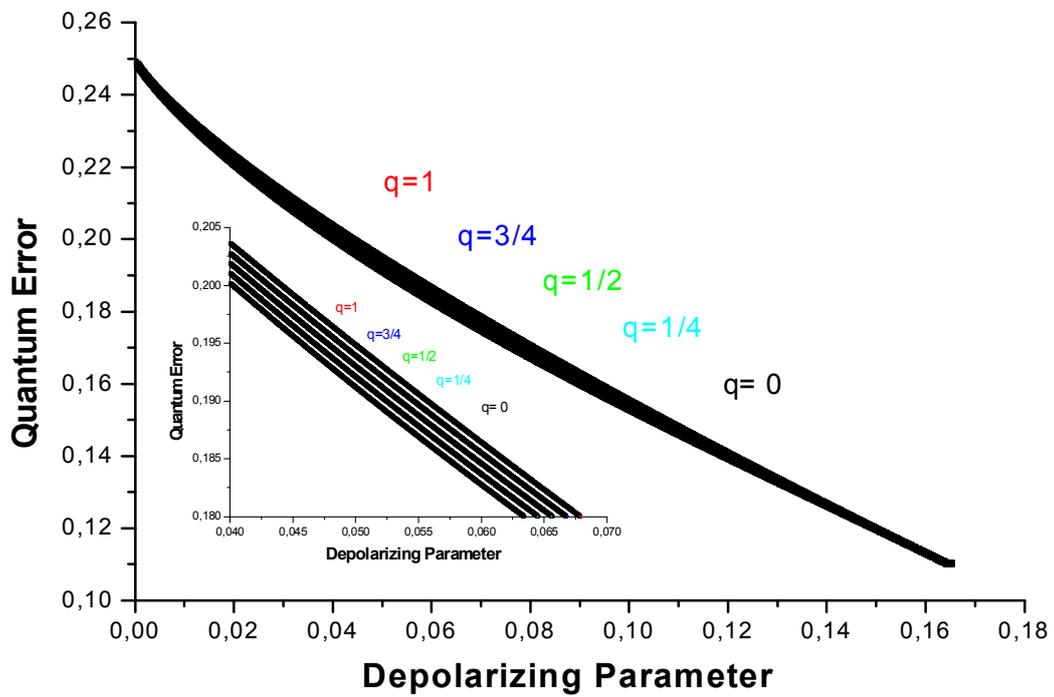

Fig.2a

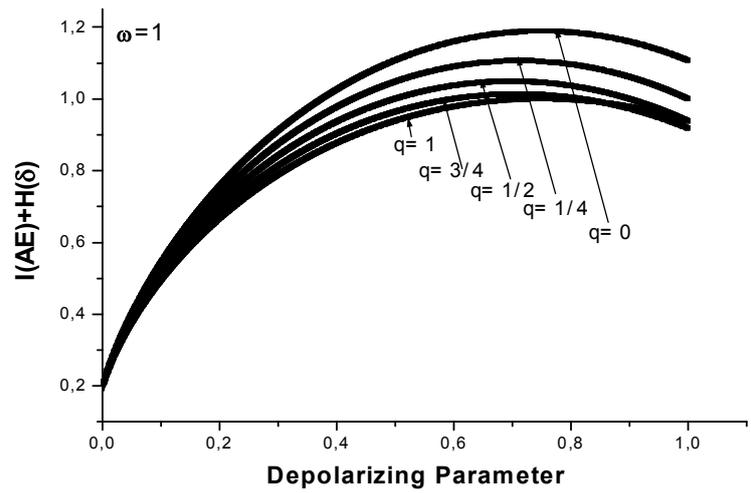

Fig.2b

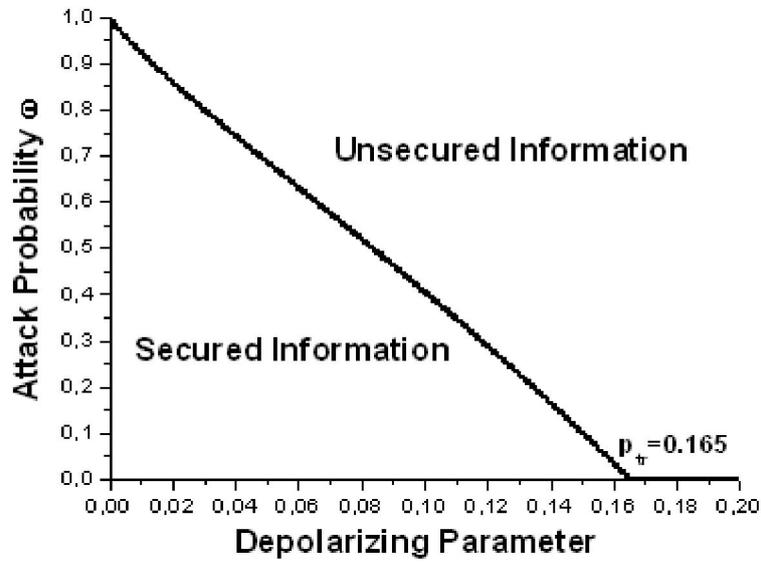

Fig.3

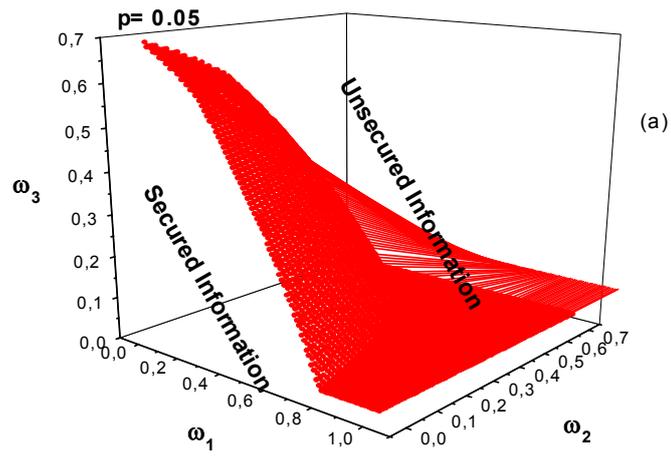

**Fig.4-a**

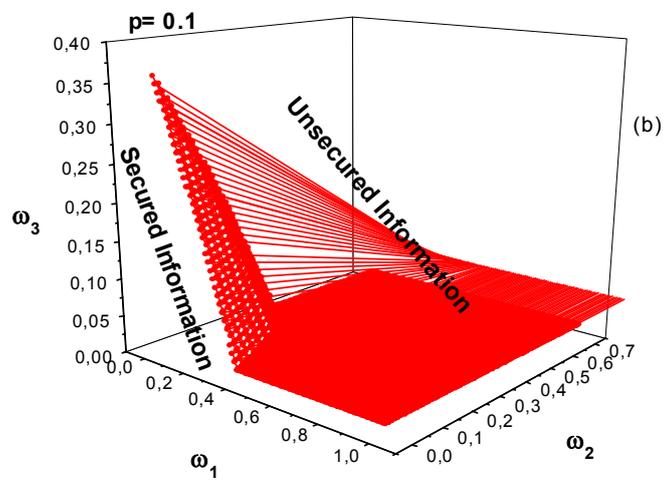

**Fig.4-b**

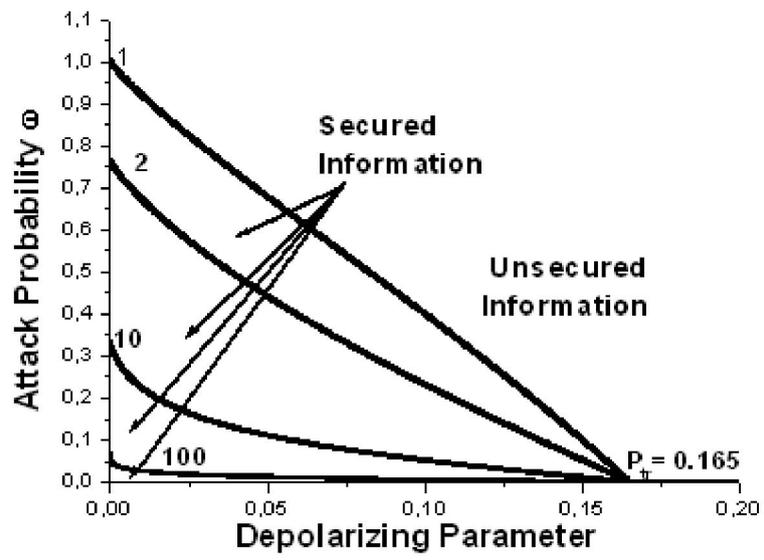

Fig.5

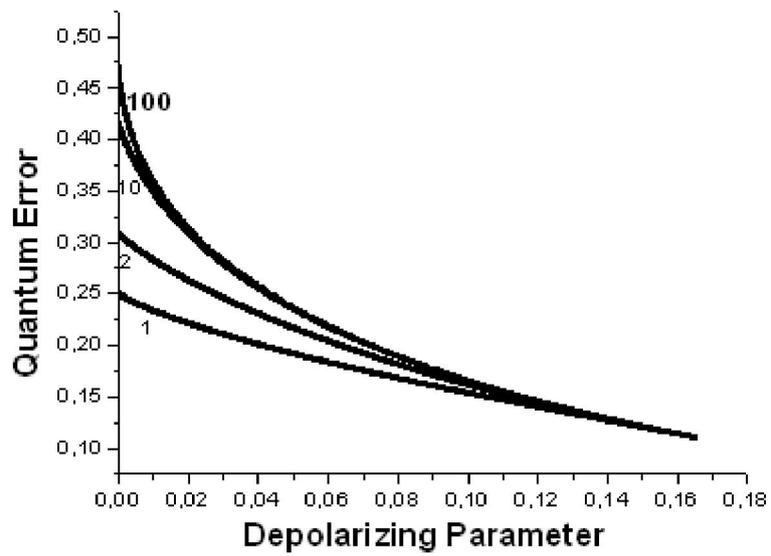

Fig.6